# User Manual of Automatic Data Curation Tool(ADCT): A bulk data curator software in Library and Information Science

Version 1.0

by
Arunavo Banerjee
Senior Project Officer (NDLI),
Master of Science, Department of Computer Science & Engineering, Indian Institute of Technology Kharagpur, India
banerjee.arunavo@yahoo.com

and

Dr. B. Sutradhar
Joint Principal Investigator (NDLI),
Librarian, Central Library,
Indian Institute of Technology Kharagpur, India
bsutra@library.iitkgp.ac.in

October 31, 2022

# Contents

# 1 Introduction

The Automatic Data Curation Tool (ADCT) is to guide and facilitate a Digital Library Curator to do batch data transformation at ease and without the exclusive need of a computer programmer. The system has two components, 1) A JSON/CSV file written in the MetaCur language for defining the curation logic, and 2) A run configuration parameter file (*.run.properties) to set up the necessary configuration parameters to run the curation logic onto required data and generate the output. The following description provides the language details and constructs to show how to perform the required tasks.

## 1.1 What is digital curation

Digital curation is maintaining and adding value to a trusted body of digital research data for current and future use; it encompasses the active management of data throughout the research life-cycle.

## 1.2 Why curate

Data are evidence supporting research and scholarship; better research is based on verifiable data, which may in turn lead to new knowledge. Observational, environmental and other data are unique and cannot be recaptured or reproduced.

Data may represent records and have associated legal requirements; curation will allow us to protect the data for the future, and manage risks.

Where data is created in the course of research that is publicly funded, a duty to manage is implied, including the provision of access to data and data reuse. Meeting this obligation will be enabled by good data stewardship.

## 1.3 What data

Curation requires effort and resources. In principle, any digital object or database may be perceived as likely to be of sufficient value for the effort and expense of curation. Data may be curated in the short-term, but may not require long-term preservation.

There are cost barriers to both curation and preservation, so an effective appraisal and selection process is essential. It is important to build an appropriate robust, distributed infrastructure to support curation.

Components may include laboratory repositories, institutional repositories, subject or discipline repositories, databases and data centres. Coordinated strategies and policies at research funder level are required, together with sufficient investment for the future.

## 1.4 When

Curation applies throughout the research life-cycle, from before or at the point of data creation, through primary use to eventual disposal. The ADCT Curation Life-cycle Model describes the processes involved in curation. The curation life-cycle may continue indefinitely and curation cannot be left to the end of primary use, for example at the end of a funded project.

## 1.5 Who should have responsibility

A number of roles and responsibilities are involved in curation and preservation practice within the Curation Life-cycle. Curation should start with the individual or group that creates or captures the information.

Curation requires a significant amount of domain knowledge; data scientists and data curators may add value to the original data. Users, custodians and re-users of the data and the funding bodies have curation responsibilities. There is currently a shortage of experienced data scientists and curators with digital preservation experience.

## 1.6 How will curation be achieved

The key is to follow good practice, including domain, national and international standards in the capture, management and archiving of data.

Processes and tools to assure easy discovery, control access and to facilitate data sharing and reuse are required. An infrastructure of data centres and trusted repositories, together with methodologies to demonstrate provenance and assure authenticity, are essential.

Curation practice in detail will depend on the domain or discipline. Data structure, scale and ownership must also be taken into account, as well as the diversity of cultures and research methodologies.

Curation can build on and fit in with current practices, for example, researchers' informal sharing of ongoing research with colleagues; or their training, or the need to comply with formal regulatory and ethical procedures.

# 2 The ADCT Language (MetaCur)

## 2.1 Overview

The ADCT language, MetaCur, follows a JSON syntax to perform various instructions related to batch data curation in a Digital Library. The philosophy behind the language is to provide the data curation instructions on each fieeld as key-value pair information. The syntax of the language is that field names are mentioned as the key and logic's to be applied to that field are enclosed in the JSON Object value. Each JSON Object Value for a particular Field Key is named as Field Translation Block (FTB). The set of instructions to be applied for a field which would undergo some data transformation or curation is to be enclosed inside a JSON Array construct 3named Action in the Field Translation Block. The Field-Field Translation Block (FTB) Pairs are then enclosed as a JSON Object within another Key Fields, inside a root JSON construct. Since the language MetaCur depends on JSON syntax and its specifics, it is good to have some idea regarding JSON format (https://www.json.org/json-en.html).

**as per A symbolic example of a logic description for a field is mentioned below:**

```
{
"Fields" : { // Fields Key. All Field-Field Translation Block Pairs are to be mentioned here.
"<fieldName1>": { // Key: The name of the field for which a data curation logic to be applied.
// Value:  Translation logic to be applied for  the field.  This is called Field
Translation Block
"<FP 1>":  <FP1 Values>, // Optional Field Property Name and Value while assignment of data.
"<FP 2>":  <FP2 Values>, // Optional Field Property Name and Value while assignment of data.
"action": [ // The container for instruction sets. Instructions are applied in the order mentioned.
"<Ins 1>", // Instruction1. Any available instruction can be applied here. */
"<Ins 2>", // Instruction2. Any available instruction can be applied here. Instruction 2 is
applied after Instruction 1.
...
],
"Ins 1": { // Instruction descriptors. Defines the instruction details if default behavior is
overriden. */
"<Ins 1 Prop 1>": "Ins 1 Prop1 Value(s)",
"<Ins 1 Prop 2>": "Ins 1 Prop 2 Value(s)",
...,
"action":[ ... ] // nested action if any
},
"Ins 2": {
...
},
...
},
"<fieldName2>":{
...
},
...
}
}
```

The instructions and the details in the above example Field Translation Block are as follows:

A Field Translation Block(FTB) constitutes three components,
1. A set of Field Properties to describe the behavior of the field during an assignment,
2. An Action block comprising a set of actions to be applied on each of the field values as per their specifications, and
3. Action Descriptors specifying details for each of the Actions in the action block.

In a FTB all of them are optional however among Field Properties and Action one group of instruction has to be present in order for the FTB to be active.

**1. Field Properties:**
Each field is characterized by a set of Field Properties. If nothing is mentioned then a default behavior is assumed for the field. In order to override the default behavior specific property and their values are to be mentioned as key value pair information within the FTB. We shall speak more about the Field Properties keys and their constructs in detail in its specific section.

**2. Action:**
The construct action as mentioned above is a JSON Array which holds the curation instructions. The instructions are applied in the order they are mentioned in the JsonArray action. Actions have their own Action descriptors and they can also be nested. We shall describe the Action and its details in its respective section.

**3. Action Descriptor:**
Each Action is described by its Action Descriptor. An Action Descriptor definition varies across actions. Filter action is common across all action descriptors. Rest of the parameters shuffles across various actions as per their usages. A common example is the parameter inputFile which is applicable across actions useMap, lookUp, and moveField since these actions require a configuration file to be associated with them. For other actions various other parameters are applied as per their behaviors. Action Descriptors are also enabled to hold nested instructions. Output of one action can be sent to another action by nesting the second action within the first. Command nesting can be done at multiple levels. This behavior is defined within the Action Descriptor of parent action by putting an Action JSONArray "action" : [<nested Action1>,<nested Action2>,...] We shall describe the details of Action Descriptors in the specific section in Action.

Let us see a Field Translation Block(FTB) below and we would dissect and explain the components step by step.

```
{ // Root Construct. Every JSON information must be enclosed within curly braces.
"Fields": { // All Field Translation logics to be provided within the Fields
```

```
Key as a JSON Object.
"dc.subject": { // A Field Translation Block(FTB) of dc.subject
"action": [ // Action JSONArray holds list of actions to be performed on
dc.subject
"lookUp", //Action lookUp to be performed on each value of dc.subject.
"copyData" //Subsequently Action copyData to be performed. ], // Types of
available Actions in would be explained in the relevant section Action
"lookUp": {  // The Action Description Block for lookUp.  "inputFile":
"lookUp.xlsx" // inputFile location to fetch the details
}
}, // Multiple FTBs can be specified separated by comma as per JSON syntax.
"dc.subject.ddc": { // A Similar FTB for dc.subject.ddc field.
"validation": true, // A Field Translation Property for dc.subject.ddc
"action": [ // A list of Actions to be performed on dc.subject.ddc
"add", // A type of Action add is performed on the field dc.subject.ddc
"copyData" // Subsequently Action copyData is performed.
], // Types of available Actions in would be explained in the relevant section
Action
"add": {
"targetValue": "091"
}
} // FTB for dc.subject.ddc ends.  List of FTBs are to be provided within
Fields Key.
} // Fields Key ends.  The FTB execution does not necessarily maintain the
writing order.
} // End of a description of data curation requirement logic.
```

Curation can be performed in two ways, 1) Collection level curation, where the logic is applied across the whole collection of data, 2) Handle_ID level curation, where individual Handle_ID can be applied by specific logic which is different within each other.

# 3  Collection Level Curation Methods

## 3.1  Run Configuration Parameters

The run configuration parameter file is a parameter file which is the second component of the ADCT system. The filename should hold an extension .run.properties. It enables the ADCT system to execute the logic file onto the source data and produce the desired output. The default assumption is that all associated files are available in the location of the run parameter file location.  If files are provided in some other location they need to be mentioned in the run configuration file as the value for configLocation parameter. There are two run modes of the ADCT system, 1. Collection level, and 2. Handle_ID level. The Collection level mode is run with the flag -c and the Handle_ID level mode is run with -h flag. The configuration file has the following parameters not required to be mentioned in any specific order, however with changing modes values of some specific parameter would change as is mentioned below. The parameters are of two types Mandatory and Optional.

A mandatory parameter is required to run the ADCT system. Without a valid
entry in them ADCT system would throw an error. Optional parameters are again
of two types, 1. With default configured, 2. Additional supplements. An
optional parameter with default configuration would be required in the system
however they need not be explicitly mentioned if the default value is not
overridden. On the other hands values of Additional Supplemental parameters
are only provided to the system if the feature is required. Availability
or Unavailability of a parameter would consequently turn on or turn off the
respective feature.

### 3.1.1 SourceData

**Usage: Mandatory**

The parameter sourceData specifies the data on which the transformation is
required to run. The sourceData can hold extensions of varied types as zipped
tar files .tar.gz of SIP files, collections or SIP Folders. Extensions are
being in progress for CSV, RIS, XLSX, BibTex files collections as well.

### 3.1.2 SourceType

**Usage: Mandatory**

The parameter sourceType helps to identify which type of source is being
processed. This is explicitly required as collections of multiple source
data can be processed in the system. Currently the supported values are
SIP-TAR, SIP-FOLDER and CSV. The extension for other formats are currently
being worked into.
SIP-TAR <Give Short Description of what is SIP TAR>
SIP-FOLDER <Give Short Description of what is SIP FOLDER>
CSV <Give Short Description of what is CSV>

### 3.1.3 TargetData

**Usage: Mandatory**

The parameter targetData specifies the output file. The current output
formats supported are zipped tar files or simple collection of SIPs.

### 3.1.4 Logic

**Usage: Mandatory**

The parameter logic holds the filename of the logic description template.
The parameter is mandatory as the basic functionality of ADCT system is
to curate according to a logic description provided by a Digital Library
Curator.
For a Collection level (-c) run the logic description template is a JSON
file. Hence any other filename extension in the logic file value apart from
.json that would throw an error from the ADCT system. Similarly for Handle_ID

level run the logic description file is required to be a CSV(Comma Separated Values) file. The logicFile parameter value for a Handle_ID level (-h) run would require to be a filename with .csv extension. The ADCT system would throw errors if the above specifications are not met.

### 3.1.5 DataReadPath

**Usage: Optional**
**Provenance: Additional Supplemental**

The parameter dataReadPath is used to mention the data path in a zipped tar file. Since a SIP-TAR may sometimes contain other configuration files generated out of the export module or being put inside to facilitate easier data transport these would put extra headache on the curator to separate those unnecessary files before processing for curation. Hence the parameter dataReadPath helps in identifying the data location inside such mixed file collections. The parameter is optional in cases where the collection contains no other files other than SIP data.

### 3.1.6 Audit Handle

**Usage: Optional**
**Provenance: Additional Supplemental**

The parameter **audit handle** specifies the **Handle_ID(s)** for which data audit reports are generated. The **audit handle** parameter enables the Data Audit module in the ADCT system. Details of this feature is discussed in Data Logging and Reporting section.

### 3.1.7 LogPath

**Usage: Optional**
**Provenance: Default Configured**

The parameter logPath mentions the path where system logs will be generated. The ADCT system generates two types of logs, 1. System logs where the curation running instance information are notified and 2. Data Audit logs, which notes the data lineage information for the document(s) as mentioned in the audit_handle parameter. The Default Configured path is the current directory from where the run configuration parameter file(*.parameter) is executed.

### 3.1.8 ConfigPath

**Usage: Optional**
**Provenance: Default Configured**

The parameter configPath mentions the path towards config files which are used in the ADCT system. Among these config files one mandatory file is the logic file. The other files include various command configuration files such as useMap.xlsx(default configuration file for useMap command), moveField.xlsx (default configuration file for moveField command) etc. The Default Configured path for configPath is the current directory from where the run configuration file is executed.

### 3.1.9 Handle_ID Format

**Usage: Optional**
**Provenance: Additional Supplemental**

Parameter **Handle_ID format** is used while curating and converting data from other formats to SIP or any other data repository ingestible formats. Since the field **Handle_ID** 5 uniquely identifies an artifact in the collective repository, creation strategies for such remains an important task.

### 3.1.10 Schema

**Usage: Mandatory**

The mandatory schema parameter plays an important role in the ADCT system. Since in the assumption of ADCT the metadata of an artifact is represented in a Key-Value pair form, nesting of keys may occur. The schema as defined in the schema parameter would also required to be a JSON file which holds this information. It simply defines the field as the key and its respective properties as JSON values of that particular key. The whole schema is represented as a single JSON object holding such multiple Key-Value pairs. The field entries in the schema file are not exhaustive, i.e., for fields which are not mentioned in the schema definition the ADCT system would not halt their operation. Instead it would proceed with a default assumption for the field behavior. Details of the schema file and field behavior are mentioned in section 3 under Field Properties sub-section.

### 3.1.11 SchemaType

**Usage: Optional**
**Provenance: Default Configured**

schemaType is a parameter specific to the application domain of the schema. It is based on the behavior of the field which might change based on their application domain however retaining the same name across all areas. The value to this parameter is specific to the application area where ADCT is being used. Combinedly schema and schemaType parameter defines the representation of a field value. Currently in the application area for NDLI the schemaType values being used are general, school and c & h. The default value for the schemaType parameter currently used is general.

### 3.1.12 ServiceIP

**Usage: Optional**
**Provenance: Additional Supplemental**

The serviceIP parameter defines the API endpoint for normalizing data representation. The API and its rendered services are coupled with the representative schema for the application area in which the ADCT system is used. In its current usage for the application domain of NDLI the data normalization service is being used via Data Service V3 API. The current version of service resides at and also used in ADCT is http://10.72.22.155:65/services/v3

### 3.1.13 CSV FieldSep

**Usage: Optional**
**Provenance: Default Configured**

### 3.1.14 CSV MultiValueSep

**Usage: Optional**
**Provenance: Default Configured**
A sample properties file would look like:-

```
sourceData=Library_of_Congress-Ph16-Export-31.12.2021.tar.gz
sourceType=SIP-TAR
targetData=Library_of_Congress-Ph16-V1.0-10.02.2021.tar.gz
logic=Phase_16_round1_logic_01.02.2022.json
dataReadPath=Library_of_Congress-Ph16-Export-31.12.2021/data
serviceIP=http://10.72.22.155:65/services/v3
```

Table 1: Collection-curation.run.properties

## 3.2 Logic File

**Field Translation Block (FTB):**

We have described the FTB in a nutshell above. We shall now describe in detail the three components of FTB and their respective principles and applications. A Field Translation Bock is the core part of the data translation description so understanding it in absolute clarity is important.

### 3.2.1 Field Properties

The Field Properties parameters combinedly describe the behavior of a field. It involves two components, i) A prior schema file definition, ii) properties key values mentioned in a FTB, however none of these are mandatory and absence of any of these would entail the default behavior specific of ADCT field parsing and translation. Hence in order to understand the complete working principle of Field Properties parameters we first need to describe the schema file and its parameters.

### Field Properties Parameters in a FTB

The Field Properties parameters which can only exclusively be defined inside a FTB are dependency and sourcePriority. The parameter dependency defines the processing dependency between two fields, i.e., the translations of one field is processed after the one it is dependent is processed, and sourcePriority which controls the priority of values assigned to a particular field. It can be used to regulate the values assigned to a single valued as well as a multi valued field in the system. The structure of dependency and sourcePriority is described below:

1. **Dependency:**

The parameter dependency is mentioned as a key in the FTB and its values are mentioned as a JSONArray. The mentioned order of fields neither defines nor order of processing

2. **SourcePriority:**

- sourcePriority on lookUp.

Requirement 1:

There requirement is replacement of handle_Id with the old data handle_Id on the new data handle_Id. So I have prepared a configuration file which is also called lookUp.

Requirement 2:

The lookUp would be run on sourcePriority, Because already there handle_Id are existing on the new data and have to replace with the old data, That is why have to be set a priority.

### 3.2.2 Schema File

The schema file contains the list of fields for which some properties would be associated. The list specifies but not necessarily prevents any field to be processed which is not mentioned. This also aids in parsing complex data as they would come in a nested key-value pair form mentioned within a data value. An example of the Field Descriptions in the schema file is shown below:

```
{
"dc.contributor.author": {
"datatype": "person",
"multiValued": true,
"controlled": false,
"validation": false
},
"dc.contributor.editor": {
"datatype": "person",
"multiValued": true,
```

```
"controlled": false,
"validation": false
},
...
...
}
```

We shall now explain the 4 parameters and their values which constitutes the description of a field in the schema, **i) datatype, ii) multiValued, iii) controlled, and iv) validation.**

- **Datatype:**

The key "datatype" categorizes the type of data which is associated with the field. Again, since the language and curation philosophy is inclusive, conformity of the field value with the type is not enforced. However if required they can be enforced by making the validation parameter true. Datatypes defined here for these fields are not only primitive, i.e., String, Integer, Date, custom data types can also be specified as per the specification and definition of the underlying schema. When validation is turned on for a field the associated conversion of the data takes place and is checked against the specifications of the datatype mentioned. This can be done by making the respective change in the schema file or can also be passed as a parameter with the same name as "**validation**" in the Field Properties Parameter section while writing the requirement. An example of such is shown below:

```
    "dc.publisher.date":{
"validation": true,
"action" : [ . . . ]
}
```

While the dc.publisher.date is mentioned as a Field in the schema file as date, making the validation properties to true mean it would ensure that the dates passed on or being translated from other fields would render it in the form dd/mm/yyyy form or any other form as agreed upon in the specification of the schema. Details of the datatypes can be found more in the respective schema specifications in use.

- **MultiValued:**

The key "multiValued" is self explanatory. It specified the existence of multi occurrences of the key in the data item. Once set as false the first occurrence of the field is being assigned during translation. Any subsequent occurrences can be controlled by using a feature called "sourcePriority" in the Field Properties Parameter. We would explain the parameter in the Field Properties Parameter section below

- **Controlled:**

The key "controlled" determines the valid values of a particular field. It is right now used as a placeholder to mark the property of the field and assignment of a valid value is left to the Digital Library Curator.

- **Validation:**

The key "validation" controls the behavior of a field. The functionality has already been described in the DataType section.

All the above Field Properties parameters can also be described in a FTB inside a logic description file.

### 3.2.3 Action Instructions:

The section Action Instructions describes the set of available actions which can be applied to the data. These actions are mentioned as a JSON Array value for the "action" key. Each action can be implemented standalone or they can be put together with other actions sequentially in order for the transformations to take effect accordingly. Actions can be either a halting action or a continuing action. A halting action means the consequent actions in the action container would not be executed after a successful execution of the current action and transformations returned from the current action are implemented to the data. However, for a continuing action the execution falls through the action container and each action which matches criteria for execution would be implemented till a halting action is encountered or the end of the container is reached. This means all the transformations during this process are collected together and are implemented to the data. Actions are associated with an action descriptor, which attributes the configurations of a specific action. If an action descriptor is not explicitly specified, default configurations are applied to that action. The following actions are currently available for MetaCur, which makes it quite exhaustive to achieve the required transformations. However there is always scope for introducing new actions and apply with the same method in the future. The set of current actions are **i) useMap, ii) copyData, iii) moveField, iv) lookUp, v) add, vi) merge, vii) curate, viii) attach and ix) filter**. We would speak about these actions and their action descriptors in details below.

### 3.2.4 useMap

The action useMap facilitates application of one to one mapping to the data. The exact key value pair as provided in the mapping details when matched with the respective field-field value pair combination of an item a transformation of the data takes place. The matching and the 9 transformation are mentioned in a configuration file which is either to be explicitly specified or configured default in its action descriptor. An exact equality between the data and the logic is verified based on the text value and the result is returned generating out of the mentioned transformation. The useMap action has two associated components, 1) A **Configuration File** which specifies the underlying transformation details and 2) An **Action Descriptor** that specifies the action parameters. In the following section we would elaborate their specifications and usages in detail.

**1.)** The Configuration File is an .xlsx file consisting of four 4 columns **A.) sourceField, B.) sourceValue, C.) targetField and D.) targetValue**. The instruction works with an exact match between the information provided in the column sourceField sourceValue pair and that with the field-field value combination of the data. All the fields for which such translations apply are included in the configuration file. Hence the configuration file becomes a single file where all such occurrences can be listed. It is extremely important to state that the column orders are static and the transformation would fail to execute if there is any change in the column order of the configuration file template of useMap, commonly known as useMap.xlsx. Also the template assumes only the first tab holds the designated data. Rest of the tabs even if present are ignored. We shall explain the columns and their applications in details below:

**1.A.) sourceField:**

The name of the field for which useMap instruction applies. This field name has to be mentioned in the logic file as a key and the action JSON Array should contain useMap in the FTB for this particular field in order for the transformation to take effect.

**1.B.) sourceValue:**

The specific value of the sourceField for which on an exact matching the requisite translation would take place. The matching is case sensitive and textual. By case sensitive, a value "AB" mentioned in the configuration file would match with only "AB" as in the data and not with any other variation as "ab" or "aB" or "Ab". By textual it is meant that numerical equality or inequality or specific properties of any other datatypes as date, URI etc., can not be applied to the data. Hence a field though it is characterized as an Integer or date or any other datatype in order to use useMap action exact field value is to be provided in the configuration file.

**1.C.) targetField:**

The column targetField is part of the transformation process as mentioned in the description of the Action. For the action useMap a transformation can be one to one or one to many. This means a key-value pair upon a successful matching would shift the value to the respective field as mentioned in the targetField column. The value to be shifted can be the original value which is the same as the sourceValue applied for equality checking or any other value as specified in the targetValue. Important thing to note here is, if such a pair requires a transformation to multiple target fields separate entries are to be made in the useMap configuration file.

**1.D.) targetValue:**

TargetValue column mentions the value to be assigned to the targetField after transformation.This can be the original value of the field, some new value or some composite value containing the original value. Since transformations can be one to one or one to many, the 10 targetValue column is enabled to hold multiple values for the respective field mentioned in the targetField column. Unlike for targetField where individual target fields need to be mentioned in separate entries for a one to many mapping scenario, multiple target values destined for a single targetField can be written in one entry separated by a single character delimiter. However this does not prevent the user from writing different target values for the same targetField, in case the delimiter can not be specified.

Below we would show an example of the useMap configuration file(useMap.xlsx) and explain each property mentioned above with an entry from that.

| sourceField | sourceValue | targetField | targetValue |
|---|---|---|---|
| lrmi.learningResourceType | researchHighlight | remove | remove |
| lrmi.learningResourceType | notes | remove | remove |
| lrmi.learningResourceType | Script | remove | remove |
| lrmi.learningResourceType | report | remove | remove |
| lrmi.learningResourceType | journal | remove | remove |
| lrmi.learningResourceType | summary | remove | remove |
| dc.description | Manuscripts. 2 p., 8 1/2 x 13, mss. (unbound) | | remove | remove |
| dc.description | Manuscripts. 10 p., 8 1/2 x 14, mss. (unbound) | | remove | remove |
| dc.contributor.author | Jackson, James & Washington | dc.contributor.author | Jackson, James;Washington |
| dc.contributor.author | Moore, [Mrs] Robert | dc.contributor.author | Moore, Robert |
| dc.contributor.author | Military Rations | ndl.sourceMeta.additionalInfo@note | Contributor Names: $value$ |
| dc.coverage.temporal | Zions Har, Boston, Massachusetts, September 28, 1892 | dc.coverage.temporal | September 28, 1892 |
| dc.coverage.temporal | Zion's Herald, Boston, Massachusetts, September | remove | remove |
| dc.coverage.temporal | Woonsocket, Rhode Island, 1982 | dc.coverage.temporal | 1982 |
| dc.coverage.temporal | WM. H. Moore, Washington, D.C., December 14, 1887 | dc.coverage.temporal | December 14, 1887 |
| dc.coverage.temporal | WM H. Alden, Philadelphia, Pennsylvania, october 3, 1894 | dc.coverage.temporal | October 3, 1894 |

Table 2: useMap

Row-1 specifies the column names in the example file and are self explanatory. The presence and order of the header row in the configuration file is extremely important as the respective column values are considered accordingly.

In the sourceField column it is important to note that the configuration file contains all those fields which apply the Action useMap. Here we can see the list of source fields contains lrmi.learningResourceType, dc.description, dc.coverage.spatial, dc.contributor.author etc. All these fields in order to apply the desired translation in the data need to specify the action useMap in the JSONArray value against the key action in their respective FTBs. Sample FTB for above example cases are mentioned below:

In the sourceField column it is important to note that the configuration file contains all those fields which apply the Action useMap. Here we can see the list of source fields contains lrmi.learningResourceType, dc.description, dc.coverage.spatial, dc.contributor.author etc. All these fields in order to apply the desired translation in the data need to specify the action useMap in the JSONArray value against the key action in their respective FTBs. Sample FTB for above example cases are mentioned below:

```
  {
"Fields": {
"lrmi.learningResourceType": {
"action": [
"useMap",
...
] },
"dc.description": {
```

```
"action": [
"useMap",
...
] },
"dc.coverage.spatial": {
"action": [
"useMap",
...
],
"useMap": {
"inputFile": "useMap.xlsx"
} }
}
}
```

It is to be noted that the configuration file when named useMap.xlsx and kept at the config location would serve as the default configuration for the useMap action. Hence it need not be separately configured in the action descriptor block and consequently if no other action configurations are specified, no action descriptor would be required for such cases as shown above for field lrmi.learningResourceType and dc.description. However they can be explicitly mentioned in the action descriptor block too as shown for the field dc.coverage.spatial for clarity if the curator feels. Separate configuration files for separate fields are not necessary, unless otherwise required. We shall explain such cases in detail in the action descriptor part where multiple useMap configuration files might be required based on filters so it helps achieve different transformations on the same field – field value pairs on different criteria.

The column sourceValue as described above contains the values to be matched. As already mentioned above the matching is exact and textual, hence a value in an Integer/ date/ other field as defined by the schema is not converted to their respective data types and then matched ith the sourceValue entry in the useMap configuration file. Row # shows an example of sourceValue for the field dc.format.extent@pageCount where the configuration entry has been put as 07 for matching with the value 07 as it has appeared for the field dc.format.extent@pageCount. Here, even if the datatype of the field is described as Integer in the respective schema JSON file ( herein, NDLIGeneralSchema.json ) the sourceValue in useMap configuration is still specified as 07 and not 7 to ensure proper  textual matching.The way to mention this in a .xlsx file varies from the software which is used to create such files. In most of the cases specifying the type of the cell as text serves the purpose otherwise starting the value with an apostrophe (') would mark the value entered in the cell as text in software's like Microsoft Excel. Similarly for fields having datatype as date matches the value in an useMap configuration file in a textual way. In Row# such examples are shown for the field dc.publisher.date. In this cases the field values are matched as they have appeared in the data and not by their contextual meaning. For example in Row# dc.publisher.date holds a value of 2021/01/03 which would exactly match the value appearing in the data in the same way, although the standard representation of the date datatype would be YYYY-MM-DD, i.e., 2021-01-03. Similarly Row# shows some more variety of such occurrences and their configuration entries in the useMap.xlsx file.

The targetField and targetValue columns are part of the transformation criteria. The column targetField would either hold a field name against which the targetValue would be assigned, or it may hold a special value called remove. The value "remove" in the target field is a reserved keyword which has the purpose of delete in the resulting transformation. If a key-value pair is matched using the configuration of useMap action then specifying remove in the targetField column would delete that particular key-value pair in the data. For any other value it is perceived as a field name and the corresponding target value is assigned as per the field properties mentioned in either Field Properties Block or in the schema information file. If a field name is not mentioned in any of these places, a field is default considered as following:

```
"<FieldName>" :
{ "datatype":"text",
"multiValued":true,
"controlled":false,
"validation":true
}
```

The final column targetValue assigns the provided value or expression to the specified targetField post transformation. The type of assignment can be of three types, delete (the matching pair from data), single value assignment, or multi value assignment. If both targetField and targetValue contains the reserved keyword "remove" the matching key-value pair from the data is deleted. Single value assignment can be of two types, static value assignment and dynamic value assignment. Both these assignments transform the matched key-value to the target forms except for dynamic assignment where target values can be substituted using a reserved keyword $value$ or $<field-name>$. If the targetValue column does not contain these keywords then the exact value mentioned is assigned to targetField after transformation. This is a static value assignment. For dynamic value assignment the notation $value$ or $<field-name>$ would be resolved to their specific references and the resultant value would be assigned accordingly to the target field post transformation. The notation $value$ refers to the current value to which the action useMap has been applied. In Row# it can be seen that if a matching occurs for the field-value pair "dc.contributor.author" and "Military Rations" a transformation takes place which assigns the value "Contributor Names: $value$" to the field "ndl.sourceMeta.additionalInfo @note". Now during the transformation $value$ is resolved to the matching value of the sourceField dc.contributor.author which is "Military Rations" and the final assignment to ndl.sourceMeta.additionalInfo@note becomes "Contributor Names: Military Rations". For $<field-name>$ substitution the respective field value from the source data is replaced in place. In case the field is not present in source data the particular notation is removed and the rest of the value is assigned to the mentioned target field. A multi value assignment is just repetition of any form of single value assignments separated by a delimiter. As is seen in Row1 targetValue column contains values separated by a delimiter ";" (semicolon) for targetField. This would make the transformation to split the mentioned entry – in the targetValue column by ";" and assign each value to the mentioned field name in the targetField column. One important thing to note here is, the delimiter can be any character based on the decision of the DL curator and need not be explicitly the ";" character. It also needs to be mentioned in the action descriptor if the delimiter has been used for a multi value transformation in the targetValue column. Otherwise the whole value is treated as a single value and necessary substitution would take place as per the process of single value assignment method defined above. The presence of a multi value transformation does not prevent any single value transformation to take effect. However, in case, if there are both of the transformations mentioned in the configuration file priority would be assigned for multi value transformation to take effect based on the delimiter mentioned in the action descriptor. This means if a transformation is intended for single valued and the multi valued delimiter is present within, it would split-ted accordingly to the delimiter and multiple values would be assigned to the mentioned target field, if supported. Otherwise the first part of the value would be assigned and the rest of the parts would be ignored if in the field properties multiValue parameter is mentioned false.

**2.) The Action descriptor for useMap contains the following:**

A) a filter action to specify the filter criteria for useMap to apply, B) an action container to hold nested actions C) parameter inputFile to identify the configuration file and D.) parameter delimiter to specify the multi value separator character. A sample useMap action descriptor looks like the following:

```
"useMap" : { "filter": [
. . . // filter specifications
],
"inputFile": "<useMap-configuration-file>",
"delimiter":"<the delimiter character>",
"action":[
. . . // nested actions
],
}
```

We would explain each parameter and their specifications below.

- **Filter:**

A filter action is an instruction specified by the key "filter". When it is mentioned in the action descriptor, the useMap configuration is applied to only those data which matches the filter criteria. A filter in useMap consists of its own specification parameters along with the other parameters of useMap. The construct for a filter when applied to the useMap command looks the following:

```
    "filter": [
"filter1",
"filter2",
...
],
"filter1": {
// filter specifications
"inputFile": "<useMap-configuration-file-for-filter1.xlsx>" // Mandatory
"delimiter":"<delimiter-character>" // Optional, only if multivalued transformation
},
"filter1": {
// filter specifications
"inputFile": "<useMap-configuration-file-for-filter2.xlsx>" // Mandatory
"delimiter":"<delimiter-character>" // Optional, only if multivalued transformation
},
...
]
```
We shall explain the filter action separately in its own section.

- **Nested Action Container:**

A Nested Action Container holds any subsequent actions after a successful application of useMap to the input data. These actions are applied to the transformed data and not on the original field-field value combination since these are nested within one another. Nesting of command can be upto any level, but the DL curator has to be cautious so as not to introduce repetition and circular transformation arising out of the nested action as the whole utility is designed towards inclusiveness of the system and does not validate the curator's decision of data management. The list of nested action is specified by the "action" key and a JSON Array construct as the value holding the action instructions in a sequential order. **InputFile:** The inputFile parameter specifies the name of the useMap configuration file. The key is "inputFile" and the value is the name along with the file extension .xlsx mentioned as a String JSON value. The default name of the configuration file is denoted as useMap.xlsx when the inputFile parameter is not specified but during the curation execution it would be prompted for verification if indeed the file is to be used. The following is the screenshot from the curation execution for inputFile configuration interaction with the user

The text in green is the user input provided during the verification step. One point to be noted is that all configuration files have to reside within the config location of run.properties configuration file. Now referring to the discussion that when it is necessary to use the inputFile parameter, multiple fields can have different delimiters for their multi value transformation configurations depending on the data and keeping in mind for the simplicity of useMap configuration addition of another column delimiter is not efficient. Hence in that particular case separate useMap configuration files can be created for the field having a different delimiter than the others. Another use of distinct input file specification is within filter action. The filter action consists of its own specification along with the parameters of useMap, i.e., based on a filter criteria certain transformations can behave differently for the same data. In those cases the useMap configuration file inside the filter action would be different than that of the configuration file for a non filtered transformation. It is to be also noted that inputFile is a mandatory parameter inside a filter action in useMap and if used then the parent configuration file becomes optional, in sense, thus facilitating the variation of application of transformation in more ways. We shall see all these examples and their applications in detail in the use cases section. **Delimiter:** The parameter delimiter is specified by using the key "delimiter" and the delimiter character(s) as a JSON String value in the action descriptor for the purpose of multi-valued transformation. The delimiter parameter can be a single character or a set of characters as applicable for the transformation defined in the target-Value column of the useMap configuration file. It has already been mentioned above that if multi valued transformations are mentioned in the configuration file but the delimiter is not specified in the action descriptor the transformations would be treated as single value transformation and the substitution rules would be applied accordingly, if applicable. It is to be noted that none of these parameters are mandatory, which means absence of an action descriptor for useMap action would not prevent the action to be applied on the data. However, in that case a default configuration to the action description is applied for the mapping and transformations to take effect. In case of a default configuration there would be no filter action, no nested action and no delimiter applicable for this action, only inputFile parameter would be default assigned to the value useMap.xlsx as a default identifier for the useMap action configuration file.

### 3.2.5  copyData

This action copies the data from one field to another based on the configuration defined in its action descriptor or to be configured default when no action descriptor is present. The action has to be specified inside an action container of a FTB in order for the copyData to take effect for that field against which the action is specified. As mentioned earlier the action copyData can be specified as a standalone action for the field or it can be put together with other actions inside the action container to be executed in an ordered way. The action copyData terminates upon a successful execution, otherwise the control falls through the rest of the actions if defined. It is a halting action. It is also important to note here that there are subtle features that make a simple copyData action interesting and non trivial. Remembering from earlier, it was mentioned that if no action has been specified for a particular field it is defaulted to copyData with targetField parameter in the action description configured to self. However, if an action container has been defined for a field and any action has been mentioned inside the container then it would be required to mention copyData command explicitly if the unmatched data is required to be copied to the desired target. Otherwise when fall through occurs in an action container absence of copyData would result in removal of the specific data.

The Action Descriptor for copyData consists of the following parameters 1) Filter Action(filter), 2) Nested Action Description(action), 3) Target Field(targetField), 4) Target Value(targetValue), and 5) Delimiter(delimiter).

The description of Filter Action and Nested Action Description has already been covered in earlier sections(<mention section ref.>) we would only focus on the copyData specific action descriptor parameters here.

- **Target Field**(targetField) :

The parameter targetField defines the field name to which the data would be copied over. In case of a default action description, where the descriptor is not explicitly mentioned, the value of targetField is the same as the source field name. However when the action description is mentioned for copyData defining the targetField name is mandatory. This scenario arises when the action descriptor for copyData contains filter action. If a filter is defined inside the copyData descriptor the absence of targetField parameter for copyData outside the filter action would signify the removal of the specific field when one or more filter matches are not found. The parameter targetField can hold a single field name or it may also hold multiple field names mentioned as a JSONArray in case of replicating the source field value to multiple target fields.

- Target Value(targetValue) :

Parameter targetValue has a similar implementation as targetField inside the action descriptor of copyData, implemented as a JSON Key. The value to the parameter is also similarly can be single valued as well as multi values which is represented using the JSONArray construct. One important token which can be used inside the target-Value parameter is *value*. The token *value* is a meta token containing a specific application for ADCT. Wherever it is used the token is replaced by the source value which is undergoing the translation. E.g., Let us say that a copyData translation is applied to a Dublin-core (dc.) field dc.creator.researcher and inside the action descriptor block the value for the parameter targetField is mentioned as dc.contributor.author and the respective value for parameter targetValue is mentioned as Prof. *value*. The outcome of this translation would result in appending Prof. token before each value of the dc.creator.researcher field and moving them to dc.contributor.author. Below is the implementation of targetValue key in the action descriptor of copyData as mentioned in the example above,

```
    "dc.creator.researcher": {
"action": [
"copyData"
],
"copyData": {
"targetField" : "dc.contributor.author"
"targetValue": "Prof. value"
}
}
```

- Delimiter(delimiter): The key

It is important to note that none of the keys in the action descriptor are mandatory, however an empty action descriptor would throw an error. The parameter keys are mutually exclusive and non

### 3.2.6 moveField

This action transforms one field-field value pair to one or multiple based on the matching criteria. The main difference between this action and the action useMap is that useMap is a very specific action and has been optimized for faster performance as it checks for an exact equality between the data and logic. It serves only one operation and is introduced based on empirical analysis. So understanding the difference between them is important for when to use which action.

Now in moveField, the action is similarly configured like useMap, i.e., it has a configuration file to specify the details of the desired operation; however the columns that formulate the logic are different from that of useMap. The configuration file for moveField consists of the following 8 columns namely sourceField, matchGroup, matchType, matchValue, targetField, transformType, targetExpr, targetReplace. In the following part we would explain the meaning and working principle of each and every configuration column.

- **sourceField:**

This column specifies the field name to which the transformation logic is to be applied. Since a configuration file is specific to the instruction rather than being specific to the field all fields which undergo the transformation are contained in the configuration file. matchGroup: The instruction moveField is enabled with all possible transformation types that can happen to a field. Hence it supports sequential as well as parallel application of logic. The column matchGroup enables the instruction to achieve that task. When a set of logics are mentioned for a sourceField moveField applies all logics to the source data in parallel, i.e., if multiple transformations exist for a value of a particular field all those are captured. However this property can be controlled using the matchGroup column. A set of logic grouped by an identifier are applied in sequence and are not available for parallel execution.

- **matchType:**

MatchType refers to a type of a match value of the value on the sourceField is called matchType. Specifically it is used in the lookUp of the configuration File.
For Example:

### 3.2.7 lookUp
### 3.2.8 add

When will be add any value to the any field, Then will be add script and their target values.
For Example:

| sourceField | match_group | src_exprType | src_expression | targetField | tgt_exprType | tgt_expression | tgt_stringValue |
|---|---|---|---|---|---|---|---|
| dc.subject.other@ccl | | matches | \d{4}-\d{2}-\d{2} | dc.publisher.date | move | | |
| ndl.subject.mesh | 1 | contains | – | dc.subject | split | – | |
| dc.coverage.temporal | | contains | ? | remove | remove | | |
| dc.title | | contains | [Untitled] | dc.title | replace | [Untitled] | |
| dc.title | | contains | (untitled) | dc.title | replace | (untitled) | |
| dc.title | | contains | , undated | dc.title | replace | , undated | |
| dc.title | | contains | Untitled and undated; | dc.title | replace | Untitled and undated; | |
| dc.title | | contains | ; Untitled; Undated | dc.title | replace | ; Untitled; Undated | |
| dc.title | | contains | ; Undated; Untitled and unidentified | dc.title | replace | ; Undated; Untitled and unidentified | |
| dc.title | | contains | Untitled, | dc.title | replace | Untitled, | |
| dc.title | | contains | ( undated ) | dc.title | replace | ( undated ) | |
| dc.title | | contains | (undated) | dc.title | replace | (undated) | |
| dc.title | | contains | undated; | dc.title | replace | undated; | |
| dc.subject | 1 | matches | .* | dc.subject | regxreplace | ,\s*$ | |
| dc.subject | 1 | count:=1 | , | dc.subject | regxreplace | (.*),(.*) | $2 $1 |
| ndl.subject.hesh | 1 | contains | – | dc.subject | split | – | |

Table 3: moveField

```
  {
"Fields": {
"dc.subject": {
"action": [
"add"
],
"add": {
"targetValue": "Post Graduate Course;Under Graduate Course;Higher Education;Remote
Studies ;Web Resources;Undergraduate/Masters Course Lectures;Undergraduate/Post
Graduate;Course e-content;Online Course Content;Undergraduate/Masters Curriculum;
UG-Course ;PG Course;Science Course;Science-related Course;UG-Course Curriculum;PG
Course Curriculum;PG Course Learning Materials;UG-Course Learning Materials;
Learning Video Course ;Course-based Video Lecture;Lecture Notes;Video;Lecture;Cours
"delimiter":";"
}
}
}
}
```

### 3.2.9  Attach

```
The action attach attaches assets to the SIP tar file data.
```

# 4  Handle_ID Level Curation Methods

```
Handle_ID level curation method is no different than Collection level curation
process with an additional process of generating the logic script by a script
generator software.
```

# 5  Curation Bundle

```
A Curation Bundle is a collection of files which are required to be submitted
to the ADCT systems. It consists of the following set of files:
```

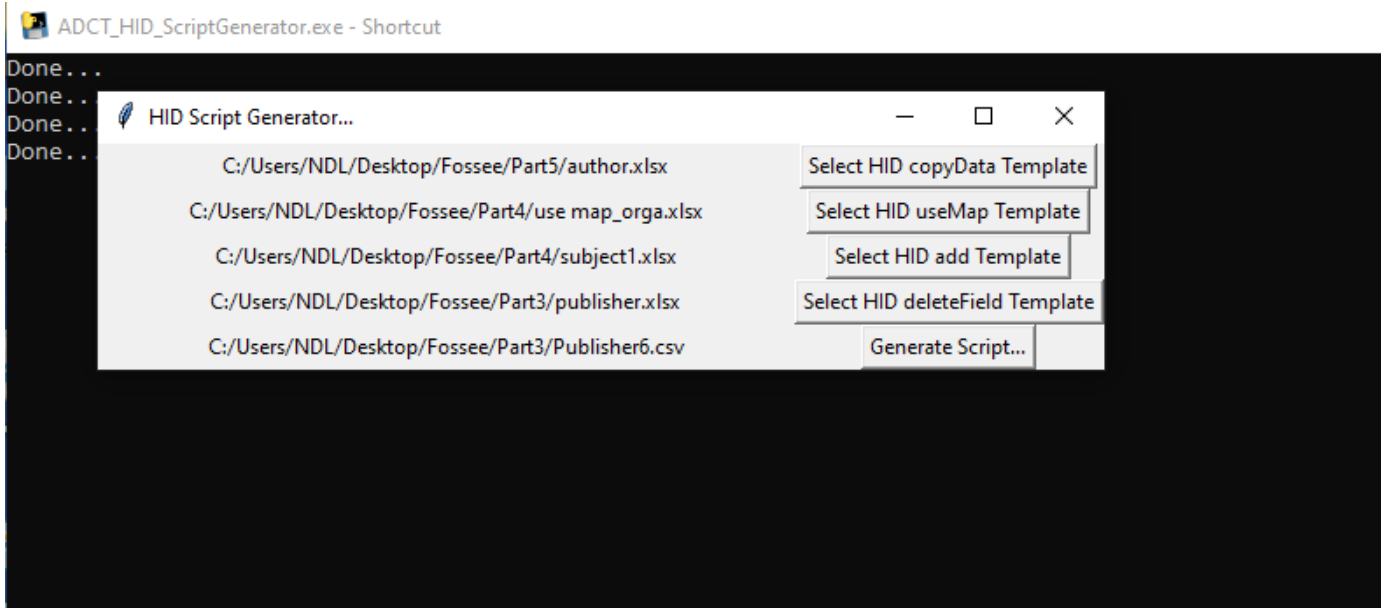

Figure 1: HID Script 1

## 5.1 Run configuration parameter files

**Provenance: File extension must be run.properties**
For Example:

    sourceData=Inflibnet-ePGP-23.08.2022.tar.gz
sourceType=SIP-TAR
targetData=Inflibnet-ePGP-V3_23.08.2022.tar.gz
logic=epgPathsalaV3_Logic.json

## 5.2 Logic File(s)

Logic File is a file where we can write all logic of the sourceField, It is
also called Logic.Json.Script. There are two works they are Field translation
block and Action Descriptors. It is write on the NotePad.
For Example:

    {
"Fields& quot":{
"ndl.sourceMeta.additionalInfo@note":{
"action":[
"lookUp",
"copyData"
],
"lookUp":{
"inputFile": "lookUp_v2.xlsx"
}
},
"Handle_ID":{
"sourcePriority"["Handle_ID","ndl.sourceMeta.additionalInfo@note"]
}

}
}

## 5.3  Action Configuration files for respective Logic File(s)

```
Action Configuration File is a File where we can write all requirement of
the Data Curation that is called Action Configuration File. Basically it is
heavy dependable on the sourceField and targetField. It is write on Excel
File.
For Example:
```

| sourceField | matchTyp | sourceValue | targetField | targetValue | targetValueType |
|---|---|---|---|---|---|
| dc.identifier.uri | equals | https://epgp.inflibnet.ac.in/epgpdata/uploads/epgp_content/S000438BE/P000725/M018751/ET/1515487623BSE_P6_M35_e-text.pdf | Handle_ID | inflibnet_epgp/business_economics_2980_E_Text | value |
| dc.identifier.uri | equals | https://epgp.inflibnet.ac.in/epgpdata/uploads/epgp_content/S000438BE/P000725/M018751/LM/1515487633BSE_P6_M35_know_more.pdf | Handle_ID | inflibnet_epgp/business_economics_2980_Learn_More | value |
| dc.identifier.uri | equals | https://www.youtube.com/watch?v=aOjUTqNazKs | Handle_ID | inflibnet_epgp/business_economics_2980_Self_learning | value |
| dc.identifier.uri | equals | https://www.youtube.com/watch?v=94NAdF5lYR8 | Handle_ID | inflibnet_epgp/economics_13391_Self_learning | value |
| dc.identifier.uri | equals | https://epgp.inflibnet.ac.in/epgpdata/uploads/epgp_content/S001827/P001855/M030300/ET/15260391699.35_ET.pdf | Handle_ID | inflibnet_epgp/hotel_amp_tourism_management_1855_30300_74427_E_Text | value |
| dc.identifier.uri | equals | https://epgp.inflibnet.ac.in/epgpdata/uploads/epgp_content/S001827/P001855/M030300/LM/15260392379.35_LM.pdf | Handle_ID | inflibnet_epgp/hotel_amp_tourism_management_1855_30300_74428_Learn_More | value |
| dc.identifier.uri | equals | https://epgp.inflibnet.ac.in/epgpdata/uploads/epgp_content/S001610/P001799/M025880/ET/1513934516Mod18Q11TrainingandDevelopment.pdf | Handle_ID | inflibnet_epgp/human_resource_management_5669_E_Text | value |
| dc.identifier.uri | equals | https://epgp.inflibnet.ac.in/epgpdata/uploads/epgp_content/S001610/P001799/M025880/LM/1513934529Mod18Q3.pdf | Handle_ID | inflibnet_epgp/human_resource_management_5669_Learn_More | value |
| dc.identifier.uri | equals | https://www.youtube.com/watch?v=M5OP5C7m_Lg | Handle_ID | inflibnet_epgp/human_resource_management_5669_Self_learning | value |
| dc.identifier.uri | equals | https://epgp.inflibnet.ac.in/epgpdata/uploads/epgp_content/S000737SP/P001448/M023360/ET/1506078970P13_M14E-text.pdf | Handle_ID | inflibnet_epgp/spanish_8485_E_Text | value |
| dc.identifier.uri | equals | https://www.youtube.com/watch?v=jC88h35c7Qw | Handle_ID | inflibnet_epgp/spanish_8485_Self_learning | value |
| dc.identifier.uri | equals | https://epgp.inflibnet.ac.in/epgpdata/uploads/epgp_content/S000737SP/P001448/M023360/LM/1506078980P13_M14KnowMore.pdf | Handle_ID | inflibnet_epgp/spanish_8485_Learn_More | value |

Table 4: Action Configuration File

## 5.4  Respective Source Data for processing the run configuration file(s).

```
sourceData is a data where ADCT Tool are run on the Data or All curation
Logic are prepare on the data that is called Respective Source Data. It may
be TAR & CSV.
For Example:
Inflibnet-ePGP-23.08.2022.tar.gz
```

## 5.5  Other custom files mentioned in the run configuration to override the default configuration mechanism.

```
For Example:
```

Figure 2: Run Configuration File

# 6 Data Logging and Reporting

Data Logging and Reporting is an essential part of any system performing an action based on user input. The ADCT system is no exception and contains an exhaustively built logging and reporting module. However both the features are optional and only enabled by user's extensive requests, i.e., they are not default enabled.

## 6.1 Data Logging

The data logging feature is called the auditing in the ADCT system. They are enabled at the run.properties configuration file to audit the data translation which has happened to specific items. The parameter which enables the Data Audit action ADCT is called **audit-handle**. For each handle a log file is generated which contains the transformations happened to the fields.

# 7 Enhancements and Bug Fixes

By the nature of its purpose the ADCT system is supposed to and has already gone through several Enhancement Requests and Bug Fixes. Since the system is at its very nascent state the Bug/Enhancement management system is not fancy, but has been integral part of the tool and sufficient in nature.

The Google Sheet has done quite well to serve the purpose. Any request for Enhancement, Change or Bug Fix needs to be entered in this Spreadsheet and the solution with respect to the request ID would be posted and updated in this section.

# 8 Appendix

## 8.1 Appendix I: ADCT Tool Demo

**This is a ADCT Tool Demo template.**

## 8.2 Appendix II: HID_useMap Template

**This is how a useMap by HID template would look like.**

| Hid | sourceField | sourceValue | targetField | targetValue |
|---|---|---|---|---|
| **12345** | dc.contributor.author | Karima Cherif | dc.contributor.author | Cherif, Karima |
| **12346** | dc.identifier.other@uniqueId | Press release No.: IFAD/03/2017 | dc.identifier.other@uniqueId | IFAD/03/2017 |
| **12347** | dc.description.abstract | This report is the most... | remove | remove |
| **12348** | dc.description.abstract | This report is the fruit ... | remove | remove |
| **12349** | dc.description.abstract | This report is the most... | remove | remove |

Table 5: Hid_useMap

# META-CUR BASIC CURATION STEP BY STEP

Mapping and Export Raw Data [WYZ_-export-09.06.2022.tar.gz]

```
                    ↙              ↘
          H.ID Based (h)        Collection Based (c)
            useMap               useMap  ("Fall Through" properties - FALSE)
            moveField            moveField ("Fall Through" properties - FALSE)
            lookup               lookup  ("Fall Through" properties - TRUE)
            copyData             copyData ("Fall Through" properties - FALSE)
```

❖ **Collection Based [-C]**

**useMap: ("Fall Through properties" - FALSE)**
https://docs.google.com/spreadsheets/d/1UajS0cSx6lc28BuDCp4wzD0w_S7L64HLCrZBxSqQ7ME/edit#gid=478213692

| sourceField | sourceValue | targetField | targetValue |
|---|---|---|---|

**moveField: ("Fall Through properties" - FALSE)**
https://docs.google.com/spreadsheets/d/1UajS0cSx6lc28BuDCp4wzD0w_S7L64HLCrZBxSqQ7ME/edit#gid=0

| sourceField | match_group | src_exprType | src_expression | targetField | tgt_exprType | tgt_expression | tgt_stringValue |
|---|---|---|---|---|---|---|---|

**Lookup: ("Fall Through" properties - TRUE)**
https://docs.google.com/spreadsheets/d/1UajS0cSx6lc28BuDCp4wzD0w_S7L64HLCrZBxSqQ7ME/edit#gid=608509111

| sourceField | matchTyp | sourceValue | targetField | targetValue | targetValueType |
|---|---|---|---|---|---|

❖ **H.ID Based [-h]**

**useMap(hid):**
https://docs.google.com/spreadsheets/d/1UajS0cSx6lc28BuDCp4wzD0w_S7L64HLCrZBxSqQ7ME/edit#gid=1323213276

| Handle_ID | sourceField | sourceValue | targetField | targetValue |
|---|---|---|---|---|
| ndl_xyz | dc.contributor.author | Karima Cherif | dc.contributor.author | Cherif, Karima |
| ndl_mno | dc.identifier.other@uniqueId | Press release No.: IFAD/03/2017 | dc.identifier.other@uniqueId | IFAD/03/2017 |
| ndl_pqr | dc.description.abstract | acd123xz | remove | remove |

Figure 3: ADCT Tool Demo

## 8.3 Appendix III: lookUp Template

**This is how a lookUp template would look like.**

**moveField(hid):**
https://docs.google.com/spreadsheets/d/1UajS0cSx6Ic28BuDCp4wzD0w_S7L64HLCrZBxSqQ7ME/edit#gid=1130133955

| Hid | sourceField | match_group | src_exprType | src_expression | targetField | tgt_exprType | tgt_expression | tgt_stringValue |
|---|---|---|---|---|---|---|---|---|

**Lookup(hid):**
https://docs.google.com/spreadsheets/d/1UajS0cSx6Ic28BuDCp4wzD0w_S7L64HLCrZBxSqQ7ME/edit#gid=456833420

| Hid | sourceField | matchTyp | sourceValue | targetField | targetValue | targetValueType |
|---|---|---|---|---|---|---|

**Add/coalesce:**
https://docs.google.com/spreadsheets/d/1UajS0cSx6Ic28BuDCp4wzD0w_S7L64HLCrZBxSqQ7ME/edit#gid=521049627

| Handle_ID | targetField | mul_sep | targetValue | mode |
|---|---|---|---|---|
| ndl_123 | lrmi.learningResourceType | | report | coalesce |
| ndl_8910 | lrmi.learningResourceType | | report | add |

**Remove_Idbased_Field:**

| Handle_ID | sourceField |
|---|---|

**Items Remove:**

| List of Handle_ID |
|---|

**\*Template:**
https://docs.google.com/spreadsheets/d/1UajS0cSx6Ic28BuDCp4wzD0w_S7L64HLCrZBxSqQ7ME/edit#gid=946672823

| 1st,.... round curation Status | Command | Round | SourceField | Logic | TargetField |
|---|---|---|---|---|---|
| | | | | | |

Figure 4: ADCT Tool Demo

# -:: At a glance Steps ::-

A. Raw Data in hand [WYZ_-export-09.06.2022.tar.gz]
B. Review and analysis
C. Prepare individual logic/curation sheet
D. Curation Sheet/ Issue Tracker with **Template***
E. Individual Logic sheet [**Configuration File**] download in excel file (.xlsx)
F. Prepare JSON script/Logic script (No script needed for HID Based Level curation)

```json
"Fields": {
    "ndl.sourceMeta.additionalInfo@RightsStatement": {
      "action": [
        "useMap",
        "copyData"
      ],
      "useMap": {
        "inputFile": "useMap_v2.0.xlsx"
      },
      "copyData": {
        "targetField": "dc.rights.holder"
      }
    },
    "ndl.sourceMeta.additionalInfo@DocumentType": {
      "action": [
        "deleteField"
      ]
    },
    "dc.description.abstract": {
      "validation": "true"
    },
    "dc.rights.license": {
      "action": [
        "useMap",
        "copyData"
      ],
      "useMap": {
        "inputFile": "useMap_v2.0.xlsx"
      }
```

G. Prepare Run Properties (xyzabc_**Curation.run.properties**)

```
sourceData=2022.Jun.08.14.04.48.2.0.tar.gz
sourceType=SIP-TAR
targetData=KlotzOnline_Stitched.V3_13.06.2022.tar.gz
logic=KlotzMath_v3_Logic.json
```

H. Prepare Bundel {1. Raw data (.tar.gz/csv/folder), 2. Logic File (JSON Script file) [For HID Level Curation Logic file will be .csv file], 3. Configuration File, 4. Run Properties File}
I. ADCT Run :-

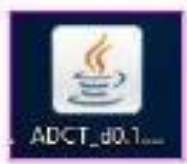

❖ Open Windows **POWER SHELL** > "Command Box" (single space) java –jar (single space) Put **Path of** ADCT_d0.2.jar **tool** (single space) - C (For

Figure 5: ADCT Tool

collection level curation) (single space) ⟶ Put *Path of* Run properties file ⟶ Press ENTER

❖ Open "Command Box" ⟶ java –jar ⟶ Put *Path of* ADCT_d0.2.jar *tool* ⟶ - h (For handle id level curation) ⟶ Put *Path of* Run properties file ⟶ Press ENTER

J. Run SynCheck tool :-

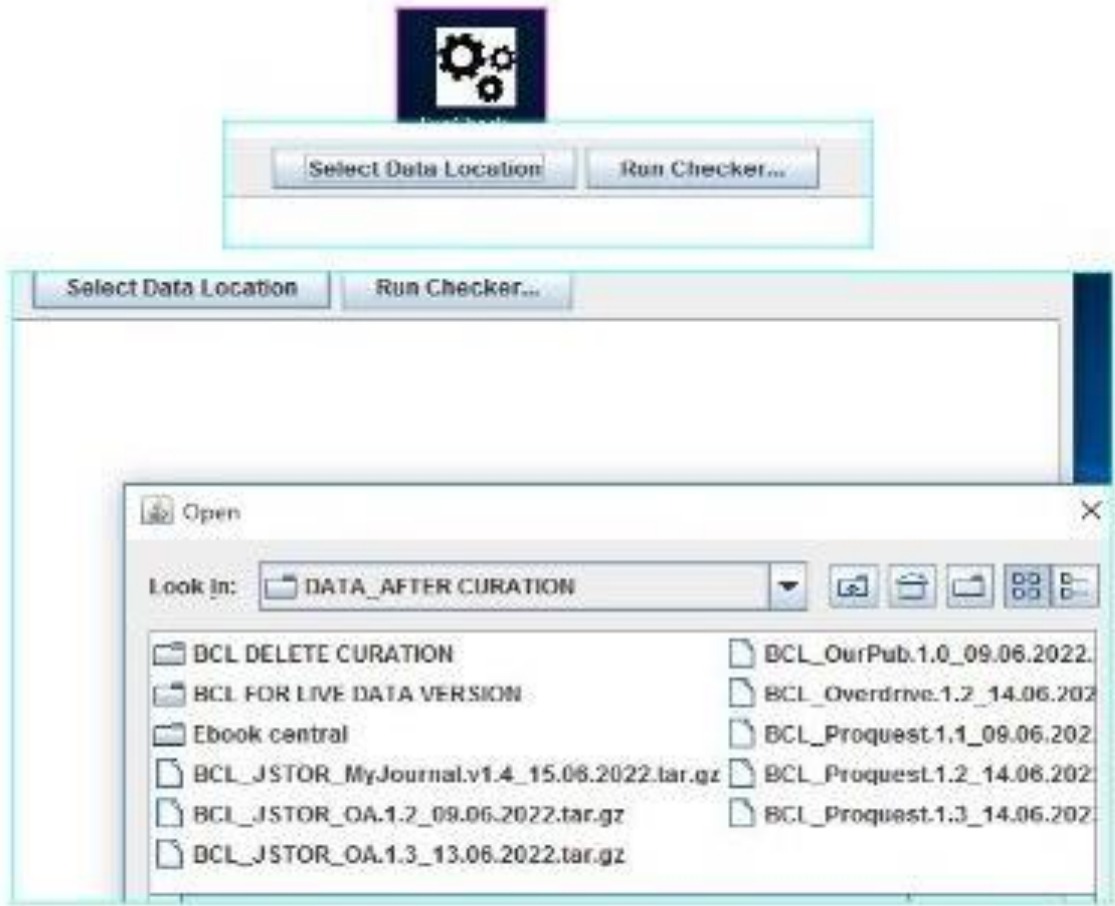

K. Ingest in 191 server of Final output data for checking
L. Ingest in NDL-TEST for QA check
M. Prepare data for Live

N.B. :- From .xlxs file (Configuration File) to .csv logic input file [For Handle ID based Curation]

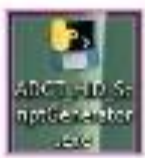 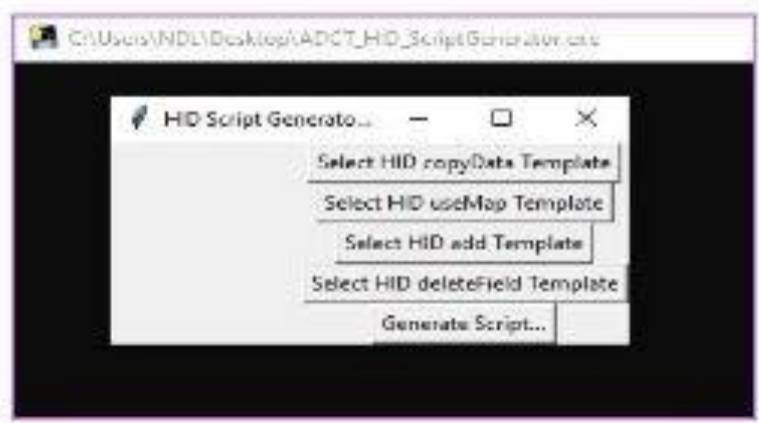

Figure 6: ADCT Tool

| sourceField | matchTyp | sourceValue | targetField | targetValue | targetValueType |
|---|---|---|---|---|---|
| **dc.language.iso** | contains | storybook | lrmi.learningResourceType | book | value |
| **dc.identifier.isbn** | contains | pdf, | dc.format.mimetype | application/pdf | value |
| **lrmi.learningResourceType** | equals | Book | dc.format.mimetype | application/pdf | value |
| **lrmi.learningResourceType** | equals | News | dc.type | text | value |

Table 6: lookup

## 8.4 Appendix IV: HID_add template

**Hid Add/coalesce**

> 1. Must be maintained Columns Order i.e. Hid>targetField>mul _sep>targetValue>Mode
> 2. Where targetValue is separated by any separator (i.e. |, ||, ; etc.), must be indicated in mul_sep column.
> 3. If there is no separator in the list, mul_sep column is blank but not deleted the column.

| Hid | targetField | mul _sep | targetValue | mode |
|---|---|---|---|---|
| 12345 | lrmi.learningResourceType | | report | coalesce |
| 12346 | lrmi.learningResourceType | | report | add |

Table 7: Hid Add/coalesce

## 8.5 Appendix V: Hid_Delete Field

**Hid_Delete FieldS**

| hid | sourceField |
|---|---|
| scilit_medknow/a42b5369079def51bc4b3f0d7c162a7a | ndl.sourceMeta.uniqueInfo@relatedContentUrl |
| scilit_medknow/f1e9e9b84ac5f64f6aed458652b561cc | dc.description.uri |
| scilit_medknow/9eb945cd46ad970195ce8f158e4ffd02 | dc.description.uri |
| scilit_medknow/5e6d24dab9dcde82a9c84eece73195c7 | dc.description.uri |
| scilit_medknow/3c4ee16c7086164ecc29b05da341d873 | dc.description.uri |
| scilit_medknow/a9c6a0d0889993ad06860bd10b0828ae | dc.publisher.place |
| scilit_medknow/a5a3a44d34121b5771ac550e4dcda232 | dc.description.uri |
| scilit_medknow/c19a4e5f10bc77df3aa102c252f12ce4 | ndl.sourceMeta.uniqueInfo@relatedContentUrl |
| scilit_medknow/c19a4e5f10bc77df3aa102c252f12ce4 | dc.description.uri |
| scilit_medknow/494d5952911584aebebfdf024a9528f8 | dc.description.uri |
| scilit_medknow/271a7d32dc7faac96da3c644ce8e13bb | dc.description.uri |
| scilit_medknow/5e6d24dab9dcde82a9c84eece73195c7 | dc.description.uri |
| scilit_medknow/3c4ee16c7086164ecc29b05da341d873 | dc.description.uri |
| scilit_medknow/cc8e155d212d86a3583731e61f55e462 | dc.publisher.place |
| scilit_medknow/f1e9e9b84ac5f64f6aed458652b561cc | dc.description.uri |
| scilit_medknow/485d06204d5c82f1387fa25a1e55c6d6 | dc.description.uri |
| scilit_medknow/6918a5e65e71ae4af307fe8d3a2a4a5b | dc.publisher.place |
| scilit_medknow/6918a5e65e71ae4af307fe8d3a2a4a5b | dc.rights.uri |
| scilit_medknow/a42b5369079def51bc4b3f0d7c162a7a | ndl.sourceMeta.uniqueInfo@relatedContentUrl |
| scilit_medknow/79c1662bd486134e3275ea0f6e830ec1 | ndl.sourceMeta.uniqueInfo@relatedContentUrl |
| scilit_medknow/b9a157f5f28239831652688d4f46b1d7 | dc.rights.uri |
| scilit_medknow/07bf97b8745b46042d010d19e866ea77 | dc.rights.uri |
| scilit_medknow/be31785820ae0e345a718474e54f3a31 | dc.description.uri |
| scilit_medknow/2a9bb9681c341c24bded57458e30954a | dc.rights.uri |
| scilit_medknow/2a9bb9681c341c24bded57458e30954a | dc.publisher.place |
| scilit_medknow/6c005f247c31534d164e7967b819893f | dc.description.uri |
| scilit_medknow/1f2dab131376355788bcd9f5287080b7 | dc.description.uri |
| scilit_medknow/2a9bb9681c341c24bded57458e30954a | dc.rights.uri |
| scilit_medknow/2a9bb9681c341c24bded57458e30954a | dc.publisher.place |
| scilit_medknow/2654efef679b29a0c802bfaae23d4cf9 | dc.description.uri |
| scilit_medknow/79c059bcdbef1644c74d66fb7592e9d6 | dc.description.uri |
| scilit_medknow/26a7a2967995a854a49aa49614684dda | dc.description.uri |
| scilit_medknow/2676bb0158cff83a5c2529f203e5d0fc | dc.description.uri |
| scilit_medknow/525166f729c623572573a310db563b41 | dc.rights.uri |
| scilit_medknow/2aac5e561681bc9809b6bbd4fd0280c6 | dc.publisher.place |
| scilit_medknow/d9dbeaa5962085f22a5e92f097ab9d73 | dc.description.uri |
| scilit_medknow/2aa08e6dfa63a04b4bce4a6b23591c8a | dc.rights.uri |
| scilit_medknow/52626e3ae26bc9e4b23069b479356e28 | dc.rights.uri |

Table 8: Hid_Delete Field